\documentclass{article}
\usepackage{fullpage}
\usepackage{hyperref}
\usepackage{here}
\usepackage{url}
\usepackage{amsmath}
\usepackage{graphicx}
\usepackage{tabularx}
\usepackage{booktabs}
\usepackage{braket}
\usepackage{algorithm}
\usepackage{algorithmic}

\newtheorem{proposition}{Proposition}
  {{\it Proof}.\ }%
  {\hfill\hbox{\rule[-2pt]{5pt}{10pt}}}

\newcommand{\Gc}{\mathcal{G}}
\newcommand{\Hc}{\mathcal{H}}

\begin{document}

\title{\bfseries Significant Subgraph Mining with\\ Multiple Testing Correction}

\author{Mahito Sugiyama\\
{\itshape The Institute of Scientific and Industrial Research, Osaka University}\\
{\ttfamily mahito@ar.sanken.osaka-u.ac.jp}
\and
Felipe Llinares L{\'o}pez\\
{\itshape Department for Biosystems Science and Engineering, ETH Z{\"u}rich}\\
{\ttfamily felipe.llinares@bsse.ethz.ch}
\and
Niklas Kasenburg\\
{\itshape Department of Computer Science, University of Copenhagen}\\
{\ttfamily niklas.kasenburg@di.ku.dk}
\and
Karsten M. Borgwardt\\
{\itshape Department for Biosystems Science and Engineering, ETH Z{\"u}rich}\\
{\ttfamily karsten.borgwardt@bsse.ethz.ch}
}

\date{}

\maketitle

\begin{abstract}
The problem of finding itemsets that are statistically significantly enriched in a class of transactions is complicated by the need to correct for multiple hypothesis testing.
 Pruning {\it untestable hypotheses} was recently proposed as a strategy for this task of significant itemset mining. It was shown to lead to greater statistical power, the discovery of more truly significant itemsets, than the standard Bonferroni correction on real-world datasets. 
An open question, however, is whether this strategy of excluding untestable hypotheses also leads to greater statistical power in {\it subgraph} mining, in which the number of hypotheses is much larger than in itemset mining.
Here we answer this question by an empirical investigation on eight popular graph benchmark datasets. We propose a new efficient search strategy, which always returns the same solution as the state-of-the-art approach and is approximately two orders of magnitude faster.
Moreover, we exploit the dependence between subgraphs by considering the {\em effective number of tests} and thereby further increase the statistical power.

\end{abstract}

\section{Introduction}
A {\em graph} is one of the most general data types to represent structured objects, and
massive amounts of structured data are now available as graphs across a wide range of domains, such as chemical compounds in PubChem~\cite{Bolton08}, biological pathways in KEGG~\cite{Kanehisa00}, protein structures in PDB~\cite{Berman00}, and social networks on the web.
Analyzing such databases, that is, {\em graph mining}, has evolved into an important branch of data mining and knowledge discovery.
Graph databases often include two or more distinct classes of graphs and, in many application domains, the ultimate purpose is to discover {\em significant subgraphs} that are statistically significantly enriched in one particular class of graphs.
In drug discovery, for instance, chemists try to identify a key substructure of chemical compounds which is significantly associated with a particular activity, e.g., anticancer activity~\cite{Takigawa13}. In a similar fashion, biologists seek substructures of proteins that are required for particular docking events~\cite{Weil09}.

Finding such significant subgraphs is an open problem, as the large number of candidate subgraphs causes both a computational and a statistical problem:
the computational problem is that it is often extremely expensive to check all subgraphs for enrichment, given that their number scales exponentially in the number of nodes of the largest graph in the database.
The statistical problem is the multiple hypothesis testing problem caused by the fact that a huge number---often billions---of subgraphs are being tested for significant enrichment, each of which represents a hypothesis. If one ignores this multiple testing problem, one may find an enormous number of {\em false positives}, subgraphs that are deemed to be significant by mistake. In particular in the natural sciences, where significant subgraphs typically undergo further experimental investigation, a large number of false positives leads to a severe waste of time and resources.
Thus {\em multiple testing correction}, calibration of the significance level in each test, is needed to control the total error rate of false positives.

Our goal in this paper is to overcome these two problems: we present {\it efficient} strategies to detect significantly enriched subgraphs {\it while correcting for multiple testing}.

A common approach to multiple testing correction is Bonferroni correction~\cite{Bonferroni36}. It tends to be highly conservative, that is, it will miss many significant observations if the number of tests performed is massive, as in graph mining or pattern mining in general. Tarone~\cite{Tarone90} proposed an improved, less conservative Bonferroni correction on categorical data. Key to this strategy is that on categorical data, only a subset of tests, called {\it testable hypotheses}, can reach significance, thereby hypotheses that are not testable can be safely removed without affecting the probability of reporting false positives. Terada et al.~\cite{Terada13} recently made it possible to enumerate testable hypotheses using a frequent itemset mining algorithm and successfully applied Tarone's insight for discovering significant combinations of transcription factors in gene regulatory network analysis.

A relevant question for graph mining is whether Tarone's strategy of only correcting for testable hypotheses can be successfully transferred to significant subgraph mining as well.
This is not a trivial question as the search space in graph mining is often exponentially larger than that in itemset mining due to combinations of vertices and edges.
In this paper, we give a positive answer to this question by
(1) extending the approach by Terada~et al.~\cite{Terada13} to solve the important open problem of significant subgraph mining with multiple testing correction via frequent subgraph mining ~\cite{Borgelt02,Inokuchi00,Nijssen04,Yan02},
(2) proposing efficient search strategies for detecting testable subgraphs, one of which is empirically orders of magnitude faster than their method, and
(3) further improving over na{\"i}ve Bonferroni correction by considering the dependence between subgraph occurrences~\cite{Moskvina08,Nyholt04}.

This paper is organized as follows:
we present our approach to significant subgraph mining in Section~\ref{sec:method}.
First we provide the necessary statistical concepts and problem statements (Sections~\ref{subsec:significant}, \ref{subsec:multhyptest}, and \ref{subsec:testable}),
then we propose search algorithms for significant subgraph detection in Section~\ref{subsec:enumeration}, followed by introducing the improved multiple testing correction via the effective number of tests in Section~\ref{subsec:effective}.
We discuss related work in Section~\ref{sec:related} and evaluate our algorithms on real-world datasets in Section~\ref{sec:experiment}.
Finally, we summarize our contributions in Section~\ref{sec:conclusion}.

\section{Method}\label{sec:method}
Let $G$ be a {\em graph}, which is mathematically defined as an ordered pair of vertices $V(G)$ and edges $E(G) \subseteq V(G) \times V(G)$.
A graph $H$ is a {\em subgraph} of $G$, denoted by $H \sqsubseteq G$, if its vertex set $V(H)$ is a subset of $V(G)$ and its edge set $E(H)$ is a subset of $E(G)$ and is restricted to its vertices, i.e., $V(H) \subseteq V(G)$ and $E(H) \subseteq (V(H) \times V(H)) \cap E(G)$.
Our notation is summarized in Table~\ref{Table:notation}.

In the following we assume that our datasets of graphs comprises two classes of graphs, but our results also transfer to more than two classes when considering one-versus-rest classification, that is enrichment of a subgraph in one class versus all others.

\begin{table}[t]
 \caption{Notation.}
 \label{Table:notation}
 \begin{center}
  \begin{tabular}{ll}
   \toprule
   $G, H$ & Graph\\
   $V(G)$ & The set of vertices of $G$\\
   $E(G)$ & The set of edges of $G$\\
   $H \sqsubseteq G$ & $H$ is a subgraph of $G$\\
   $\Gc, \Gc'$ & A set of graphs\\
   $\Hc$ & The set of subgraphs in $\Gc \cup \Gc'$: $\Hc = \{H \sqsubseteq G \mid G \in \Gc \cup \Gc'\}$\\
   $|X|$ & Cardinality of $X$\\
   $n$ (resp. $n'$) & Cardinality of $\Gc$ (resp. $\Gc'$): $n = |\Gc|$ and $n' = |\Gc'|$\\
   $x$ (resp. $x'$) & Frequency of $H$ in $\Gc$ (resp. $\Gc'$): $x = |\set{G \in \Gc | H \sqsubseteq G}|$\\
   $q(x)$ & Probability $\binom{n}{x}\binom{n'}{x'} / \binom{n + n'}{x + x'}$\\
   $f(H)$ & Frequency of $H$ in $\Gc \cup \Gc'$: $f(H) = x + x' = |\set{G \in \Gc \cup \Gc' | H \sqsubseteq G}|$\\
   $\sigma$ & Frequency\\
   $\psi(\sigma)$ & Minimum $P$ value of frequency $\sigma$: $\psi(\sigma) = \binom{n}{\sigma} / \binom{n + n'}{\sigma}$\\
   $\Hc$ & The set of subgraphs in $\Gc \cup \Gc'$, $|\Hc|$ is the Bonferroni correction factor\\
   $\alpha$ & Significance level\\
   $k$ & Natural number\\
   $m(k)$ & The value $|\set{H \in \Hc | \psi \circ f(H) \le \alpha / k}|$\\
   $k_{\mathrm{rt}}$ & (Rounded) Root of $m(k) - k$: $m(k_{\mathrm{rt}} - 1) > k_{\mathrm{rt}} - 1$, $m(k_{\mathrm{rt}}) \le k_{\mathrm{rt}}$\\
   $\tau(\Hc)$ & The set of testable subgraphs: $\tau(\Hc) = \set{H \in \Hc | \psi \circ f(H) \le \alpha / k_{\mathrm{rt}}}$\\
   $\sigma_{\mathrm{rt}}$ & (Rounded) Root frequency such that\\
   & $|\Set{H \in \Hc | f(H) \ge (\sigma_{\mathrm{rt}} - 1)}| > \alpha / \psi(\sigma_{\mathrm{rt}} - 1)$ and\\
   & $|\Set{H \in \Hc | f(H) \ge \sigma_{\mathrm{rt}}}| \le \alpha / \psi(\sigma_{\mathrm{rt}})$\\
   $\sigma_{\min}$ & The minimum possible frequency $\sigma_{\min}$ satisfying $\psi(\sigma_{\min}) < \alpha$\\
   $\sigma_{\max}$ & The maximum possible frequency $n$\\
   $s(\Hc)$ & The set of significant subgraphs\\
   $m_{\text{eff}}$ & The effective number of tests within the testable subgraphs\\
   \bottomrule
  \end{tabular}
 \end{center}
\end{table}

\subsection{Statistically significant subgraphs}\label{subsec:significant}
Suppose we are given two collections of graphs $\Gc$ and $\Gc'$, where the numbers of graphs in these sets are $|\Gc| = n$ and $|\Gc'| = n'$ with $n \le n'$ without loss of generality.
For each subgraph $H \sqsubseteq G$ with $G \in \Gc \mathop{\cup} \Gc'$, we formulate a \emph{null hypothesis} that the occurrence of the subgraph $H$ is independent from the class membership of $G$. Our task is to find for which subgraphs $H$ the data provide enough evidence to \emph{reject} the null hypothesis and to deem $H$ as a significant subgraph associated with the class membership.

From given data, we measure the statistical association between two binary random variables: the indicator vector of the class membership and the occurrence/absence of the subgraph $H$ within each graph $G$ in the database.

Let $x$ and $x'$ be the frequencies of $H$ in $\Gc$ and $\Gc'$, respectively. That is,
$x = |\set{G \in \Gc | H \sqsubseteq G}|$ and $x' = |\set{G \in \Gc' | H \sqsubseteq G}|$, as represented in the following $2 \times 2$ contingency table.

\begin{center}
 \begin{tabular}{cccc}
  \toprule
         & Occurrences & Non-occurrences & Total\\ \midrule
  $\Gc\phantom{'}$  & $x\phantom{'}$ & $n - x$ & $n\phantom{'}$\\
  $\Gc'$ & $x'$ & $n' - x'$ & $n'$\\
  Total  & $x + x'$ & $(n - x) + (n' - x')$ & $n + n'$\\
  \bottomrule
 \end{tabular}
\end{center}

The strength of the association between binary random variables is quantified as a {\em $p$-value}, defined as the probability of observing an association at least as strong as the one present in the data under the assumption that the null hypothesis of independence holds true. 
To compute the $p$-value, {\em Fisher's exact test} is commonly used. It relies on the fact that, when the margins $x +x'$, $n$, and $n+n'$ are fixed, the probability $q(x)$ of obtaining these counts $x$ and $x'$ is given by the hypergeometric distribution:
\[
 q(x) = \binom{n}{x} \binom{n'}{x'} \Bigg/ \binom{n + n'}{x + x'}.
\]
Formally, define $P_{\mathrm{L}}$ and $P_{\mathrm{R}}$ as the left-tail and the right-tail of the hypergeometric distribution, respectively. That is, given an observed count $x$, $P_{\mathrm{L}}$ is the probability of observing a smaller count 
and $P_{\mathrm{R}}$ the probability of observing a larger one:
\[
 P_{\mathrm{L}} = \sum_{X = \max\{0, x + x' - n'\}}^{x} \,q(X),\hspace*{10pt}
 P_{\mathrm{R}} = \sum_{X = x}^{\min\{x + x', n\}} \,q(X),
\]
They are used as one-tailed $p$-values, and a two-tailed $p$-value $P_{\mathrm{D}}$ is defined as\footnote{We can choose other definitions for a two-tailed test, e.g., summing up all probabilities that are smaller than $q(x)$. The analysis in this paper still holds with minor modifications.} $P_{\mathrm{D}} = 2 \min\{P_{\mathrm{L}}, P_{\mathrm{R}}\}$~\cite{Bland00}.

We say that a subgraph $H$ is {\em statistically significant} if its $p$-value is smaller than a predetermined significance level $\alpha$. Note that, by construction, $\alpha$ equals to  the {\em Type I error probability}; the probability of falsely deeming a subgraph significant.

\subsection{Multiple hypothesis testing}\label{subsec:multhyptest}
In our setup, one must test {\em all} subgraphs in a database.
The procedure described above guarantees that, for a single subgraph, the probability of being a false positive is upper bounded by $\alpha$. However, when many hypotheses are tested in parallel, the probability that at least one subgraph is a false positive, called the {\em Family-Wise Error Rate} (FWER), approaches one. This is the well-known {\em multiple hypothesis testing} problem.

To deal with this issue, one needs to {\em correct} the significance level $\alpha$ in each test to guarantee that $\mathrm{FWER} \le \alpha$. The most common method is the {\em Bonferroni correction}~\cite{Bonferroni36}, which simply divides $\alpha$ by the number $m$ of tests. The resulting FWER can be readily shown to be smaller than $\alpha$.
The number of tests $m$ is called the {\em Bonferroni factor}, which in our case is the same as the number of subgraphs.
Despite its popularity, the Bonferroni correction is known to be too conservative in many cases, that is, the statistical power, the probability to detect truly significant subgraphs, becomes too small. The problem is even more extreme in our application: as $m$ is the huge number of subgraphs tested, the Bonferroni corrected significance level $\alpha / m$ is so small that hardly any subgraph can ever reach the significance level.

\subsection{Testable subgraphs}\label{subsec:testable}
Tarone~\cite{Tarone90} showed that when testing the association of discrete random variables, as in our setup, one can improve the Bonferroni correction. The key idea is that the discreteness of the problem implies the existence of a minimum achievable $p$-value for each subgraph $H$. Let $f(H) = |\Set{G \in \Gc \cup \Gc' | H \sqsubseteq G}|=x+x'$ be the frequency of $H$ in the whole set of graphs $\Gc \cup \Gc'$, and
assume that $f(H) \le n$. 

If the marginals $f(H)$, $n$, and $n'$ are fixed, the minimum $p$-value, denoted by $\psi(f(H)) = \psi \circ f(H)$, is achieved for the most biased case when $x = 0$ or $x=f(H)$.
Since $P_{\mathrm{L}}$ and $P_{\mathrm{R}}$ are minimized at $x = \max\{0, f(H) - n'\}$ and $x = \min\{f(H) , n\}$,
their minimum values are $q(0)$ and $q(f(H))$, respectively.
From $n \le n'$, $q(f(H)) \le q(0)$ holds.
Thus we have
\[
 \psi \circ f(H) = q(f(H)) = \binom{n}{f(H)} \Bigg/ \binom{n + n'}{f(H)}
\]
for a one-tailed test, and this value is doubled for a two-tailed test.
If $f(H) > n$ and hence $f(H) = x + x' > (n + n') / 2$, we follow the definition in \cite[Supporting Text 4]{Terada13}, that is,
we simply define $\psi \circ f(H) = 1 / \binom{n + n'}{n}$.
Then $\psi$ is always monotonically decreasing, which is required for our algorithms.

If the minimum $p$-value $\psi \circ f(H)$ is larger than the significance threshold,
the subgraph $H$ can never be significant regardless of the class membership of the graphs in which it occurs. Tarone's insight is that such {\em untestable} subgraphs do not increase the FWER, and hence we can exclude them from candidate subgraphs and reduce the Bonferroni factor.
Formally, let $\Hc = \{\,H \sqsubseteq G \mid G \in \Gc \cup \Gc'\,\}$ be the set of all subgraphs in the database and define for each natural number $k$
\[
  m(k) = |\Set{H \in \Hc | \psi \circ f(H) \le \alpha / k}|,
\]
as the number of subgraphs whose minimum achievable $p$-value is smaller than $\alpha / k$.
Let $k_{\mathrm{rt}}$ satisfy
\[
 m(k_{\mathrm{rt}} - 1) > k_{\mathrm{rt}} - 1\ \text{and}\ m(k_{\mathrm{rt}}) \le k_{\mathrm{rt}},
\]
that is, $k_{\mathrm{rt}}$ is the rounded {\em root} of $m(k) - k$.
Since $m(k)$ monotonically decreases as $k$ increases, we have $m(k) - k > 0$ for all $k < k_{\mathrm{rt}}$ and $m(k) - k \le 0$ for all $k \ge k_{\mathrm{rt}}$.
Then we can see that $\mathrm{FWER} \le \alpha$ even if we reduce the Bonferroni factor from $|\Hc|$ to $m(k_{\mathrm{rt}})$, since we have
\begin{flalign*}
 \mathrm{FWER} &\le \sum \Set{\psi \circ f(H) | \psi \circ f(H) \le \alpha / k_{\mathrm{rt}},\ H \in \Hc}\\
               &\le m(k_{\mathrm{rt}}) \frac{\alpha}{k_{\mathrm{rt}}} \le \alpha.
\end{flalign*}
As a result, we have the set of {\em testable subgraphs} $\tau(\Hc)$, which is given by
\[
 \tau(\Hc) = \Set{H \in \Hc | \psi \circ f(H) \le \alpha / k_{\mathrm{rt}}},
\]
and our task of detecting all significant subgraphs is achieved by finding the root $k_{\mathrm{rt}}$ and enumerating the set $\tau(\Hc)$ of testable subgraphs.

Terada et al.~\cite{Terada13} used Tarone's method in the context of itemset mining for discovering gene regulatory motifs, where efficient enumeration of testable itemsets was achieved by applying a frequent itemset mining algorithm.
Next we show how to apply Tarone's method to significant subgraph mining.

\subsection{Enumeration of testable subgraphs}\label{subsec:enumeration}
To use Tarone's results for our purpose, the challenge is now to efficiently compute all testable hypotheses, that is all testable subgraphs.
Here we show how to use {\em frequent subgraph mining} to enumerate all testable subgraphs.
Frequent subgraph mining algorithms find all subgraphs whose frequencies are higher than the user specified threshold $\sigma$ (or its ratio $\theta = \sigma / (n + n')$).
Since the minimum $p$-value $\psi$ is a monotonically decreasing function (the proof is provided in~\cite[Supporting Text 4]{Terada13}), we have $\psi \circ f(H) \le \psi(\sigma)$ for every frequent subgraph $H$.
\begin{proposition}
 The set of testable subgraphs $\tau(\Hc)$ coincides with the set of frequent subgraphs for the threshold $\sigma_{\mathrm{rt}}$ such that
 \begin{flalign*}
 &|\Set{H \in \Hc | f(H) \ge (\sigma_{\mathrm{rt}} - 1)}| > \alpha / \psi(\sigma_{\mathrm{rt}} - 1),\\
 &|\Set{H \in \Hc | f(H) \ge \sigma_{\mathrm{rt}}}| \le \alpha / \psi(\sigma_{\mathrm{rt}})
 \end{flalign*}
\end{proposition}
{\it Proof}.\ We have for $k_{\mathrm{rt}} = \alpha / \psi(\sigma_{\mathrm{rt}})$,
\begin{flalign*}
 m(k_{\mathrm{rt}}) = m(\alpha / \psi(\sigma_{\mathrm{rt}}))
 &= |\Set{H \in \Hc | \psi \circ f(H) \le \alpha / k_{\mathrm{rt}}}|\\
 &= |\Set{H \in \Hc | \psi \circ f(H) \le \psi(\sigma_{\mathrm{rt}})}|\\
 &= |\Set{H \in \Hc | f(H) \ge \sigma_{\mathrm{rt}}}|. &\hfill\hbox{\rule[-2pt]{5pt}{10pt}}
\end{flalign*}

In the following, we present four variants to efficiently find this (rounded) {\em root frequency} $\sigma_{\mathrm{rt}}$ and enumerate all testable subgraphs.
Note that every single method gives exactly the same root frequency and testable subgraphs, resulting in the same significant subgraphs. Our search procedures can be combined with any of the many algorithms for frequent subgraph mining (an FSM algorithm for short), e.g., with AGM~\cite{Inokuchi00}, gSpan~\cite{Yan02}, Mofa~\cite{Borgelt02}, or Gaston~\cite{Nijssen04}, as long as they report actual frequencies of detected frequent subgraphs.

An important property of our search algorithms is that they require a significance level $\alpha$ as an input
but do not require the frequency threshold to be prespecified, which is attractive as it is often difficult to find an appropriate frequency threshold for a particular problem in practice.

\subsubsection*{One-pass search}
The first method is to apply an FSM algorithm only once to get the full spectrum of subgraphs (Algorithm~\ref{Algorithm:onepass}).
Since the root frequency should satisfy $\psi(\sigma_{\mathrm{rt}}) < \alpha$, we can compute the minimum possible frequency $\sigma_{\min}$ satisfying $\psi(\sigma_{\min}) < \alpha$ from $n$ and $n'$ in advance.
Then we run an FSM algorithm with this frequency $\sigma_{\min}$.
The mining process might be expensive since this $\sigma_{\min}$ is usually small, resulting in an exponentially large number of frequent subgraphs that may include many untestable subgraphs.
But once we finish mining and obtain the actual frequency $f(H)$ for all detected frequent subgraphs $H$,
we can easily obtain the root frequency, for example, by sorting the subgraphs according to their frequencies and checking them one by one, starting with the smallest frequency.

\begin{algorithm}[t]
 \caption{One-pass search}
 \label{Algorithm:onepass}
 \begin{algorithmic}
  \STATE {\bfseries Input:} Datasets $\Gc$, $\Gc'$ and significance level $\alpha$
  \STATE {\bfseries Output:} All significant subgraphs
  \STATE $\sigma_{\min} \gets 1$
  \WHILE{$\psi(\sigma_{\min}) > \alpha$}
  \STATE $\sigma_{\min} \gets \sigma_{\min} + 1$
  \ENDWHILE
  \STATE \COMMENT{$\sigma_{\min}$ is the minimum possible frequency}
  \STATE $\Hc(\sigma_{\min}) \gets \{H \in \Hc \mid f(H) \ge \sigma_{\min}\}$
  \STATE \COMMENT{This set is obtained by running an FSM}
  \STATE \COMMENT{algorithm with the threshold $\sigma_{\min}$}
  \STATE $\sigma_{\mathrm{rt}} \gets \sigma_{\min}$
  \WHILE{$|\set{H \in \Hc | f(H) \ge \sigma_{\mathrm{rt}}} | > \alpha / \psi(\sigma_{\mathrm{rt}})$}
  \STATE $\sigma_{\mathrm{rt}} \gets \sigma_{\mathrm{rt}} + 1$
  \ENDWHILE
  \STATE \COMMENT{$\sigma_{\mathrm{rt}}$ is the root frequency}
  \STATE $\tau(\Hc) \gets \set{H \in \Hc | f(H) \ge \sigma_{\mathrm rt}}$
  \STATE \COMMENT{Testable hypotheses}
  \STATE $s(\Hc) \gets \set{H \in \tau(\Hc) | P\ \text{value of}\ H\ < \alpha / |\tau(\Hc)|}$
  \STATE Output $s(\Hc)$
 \end{algorithmic}
\end{algorithm}

\subsubsection*{Decremental search (LAMP search)}
The second approach is to decrease the frequency from the maximum possible value until reaching the root frequency, proposed in LAMP by Terada et al.~\cite{Terada13} to find testable itemsets (Algorithm~\ref{Algorithm:decremental}).
We start from the maximum possible frequency $\sigma_{\max} = n$ and repeatedly run an FSM algorithm while decreasing the threshold $\sigma$ one by one as long as the condition $|\set{H \in \Hc | f(H) \ge \sigma}| \le \alpha / \psi(\sigma)$ is satisfied.
Otherwise if we have $|\set{H \in \Hc | f(H) \ge \sigma}| > \alpha / \psi(\sigma)$
at some frequency $\sigma$, the root $\sigma_{\mathrm{rt}} = \sigma + 1$.
This search is expected to be more efficient then the above one-pass search since mining with high frequency is usually much cheaper than that with low frequency and
we do not need to run the FSM algorithm with a frequency threshold lower than $\sigma_{\mathrm{rt}} - 1$.

\begin{algorithm}[t]
 \caption{Decremental search (LAMP search)}
 \label{Algorithm:decremental}
 \begin{algorithmic}
  \STATE {\bfseries Input:} Datasets $\Gc$, $\Gc'$ and significance level $\alpha$
  \STATE {\bfseries Output:} All significant subgraphs
  \STATE $\sigma_{\mathrm{rt}} \gets n$ \hspace*{10pt}\COMMENT{the maximum possible frequency}
  \REPEAT
  \STATE $\Hc(\sigma_{\mathrm{rt}}) \gets \{H \in \Hc \mid f(H) \ge \sigma_{\mathrm{rt}}\}$
  \STATE \COMMENT{This set is obtained by running an FSM}
  \STATE \COMMENT{algorithm with the threshold $\sigma_{\mathrm{rt}}$}
  \STATE $\sigma_{\mathrm{rt}} \gets \sigma_{\mathrm{rt}} - 1$
  \UNTIL{$|\Hc(\sigma_{\mathrm{rt}})| > \alpha / \psi(\sigma_{\mathrm{rt}})$}
  \STATE $\sigma_{\mathrm{rt}} \gets \sigma_{\mathrm{rt}} + 2$ \hspace*{10pt}\COMMENT{$\sigma_{\mathrm{rt}}$ is the root frequency}
  \STATE $\tau(\Hc) \gets \set{H \in \Hc | f(H) \ge \sigma_{\mathrm rt}}$
  \STATE \COMMENT{Testable hypotheses}
  \STATE $s(\Hc) \gets \set{H \in \tau(\Hc) | P\ \text{value of}\ H\ < \alpha / |\tau(\Hc)|}$
  \STATE Output $s(\Hc)$
 \end{algorithmic}
\end{algorithm}

\subsubsection*{Incremental search}
Instead of decreasing the frequency, here we newly propose the opposite strategy, that is, increasing the frequency one by one (Algorithm~\ref{Algorithm:incremental}).
We use an additional trick, {\em early termination} of an FSM algorithm for frequencies $\sigma < \sigma_{\mathrm{rt}}$.
For such a frequency $\sigma$, we know in advance that the number of admissible subgraphs at this frequency is at most $\alpha / \psi(\sigma)$ --- if it is larger, $\sigma$ cannot be the root frequency.
Thus during the process of subgraph mining, we are able to terminate it as soon as the number of subgraphs exceeds this value.
The whole process is as follows:
we start from the minimum possible frequency $\sigma_{\min}$ and repeatedly apply an FSM algorithm while increasing the threshold $\sigma$ one by one, as long as the search process terminates early.
Otherwise if mining is finished at some frequency, this frequency is the root.
This approach is also expected to work efficiently, as the number of admissible subgraphs is quite small if the frequency $\sigma$ is small.
Therefore we can quickly increase the frequency and, moreover,
we have to finish the full mining process only once (i.e., without early termination), for the frequency $\sigma_{\mathrm{rt}}$.
Thus the complexity is the same as an FSM algorithm itself.

In parallel to our work, a sped up version of LAMP was published by Minato et al.~\cite{Minato14} for significant itemset mining, which also uses incremental search.
Unlike our approach, in which pattern mining and incremental search can be combined in an arbitrary, modular fashion, they change the mining process itself to prune untestable hypotheses as early as possible.

\begin{algorithm}[t]
 \caption{Incremental search}
 \label{Algorithm:incremental}
 \begin{algorithmic}
  \STATE {\bfseries Input:} Datasets $\Gc$, $\Gc'$ and significance level $\alpha$
  \STATE {\bfseries Output:} All significant subgraphs
  \STATE $\sigma_{\mathrm{rt}} \gets 1$
  \WHILE{$\psi(\sigma_{\mathrm{rt}}) > \alpha$}
  \STATE $\sigma_{\mathrm{rt}} \gets \sigma_{\mathrm{rt}} + 1$
  \ENDWHILE
  \STATE \COMMENT{This is the minimum possible frequency}
  \REPEAT
  \STATE Run an FSM algorithm with the threshold $\sigma_{\mathrm{rt}}$
  \STATE with monitoring the number $m$ of frequent subgraphs
  \IF{$m > \alpha / \psi(\sigma_{\mathrm{rt}})$ while the process}
  \STATE Terminate the mining process
  \ELSE
  \STATE $\Hc(\sigma_{\mathrm{rt}}) \gets \{H \in \Hc \mid f(H) \ge \sigma_{\mathrm{rt}}\}$
  \STATE \COMMENT{This set is obtained by running an FSM}
  \STATE \COMMENT{algorithm with the threshold $\sigma_{\mathrm{rt}}$}
  \ENDIF
  \STATE $\sigma_{\mathrm{rt}} \gets \sigma_{\mathrm{rt}} + 1$
  \UNTIL{the mining process is not terminated}
  \STATE $\sigma_{\mathrm{rt}} \gets \sigma_{\mathrm{rt}} - 1$ \hspace*{10pt}\COMMENT{$\sigma_{\mathrm{rt}}$ is the root frequency}
  \STATE $\tau(\Hc) \gets \set{H \in \Hc | f(H) \ge \sigma_{\mathrm rt}}$
  \STATE \COMMENT{Testable hypotheses}
  \STATE $s(\Hc) \gets \set{H \in \tau(\Hc) | P\ \text{value of}\ H\ < \alpha / |\tau(\Hc)|}$
  \STATE Output $s(\Hc)$
 \end{algorithmic}
\end{algorithm}

\subsubsection*{Bisection search (LEAP search)}
Since our task can be viewed as a root-finding problem, we can apply the well-known {\em bisection method} as our fourth approach (Algorithm~\ref{Algorithm:bisection}).
This strategy is used in LEAP by Yan et al.~\cite{Yan08} to obtain top-$k$ subgraphs in terms of a user-specified objective function in which a statistical test can be used, yet without multiple testing correction.
Thereby we exploit only its search strategy to find the root frequency.
It repeatedly bisects an interval of possible frequencies and selects a subinterval in which the root frequency lies.
First we set the interval $[a, b]$ from the minimum possible frequency $a = \sigma_{\min}$ to the maximum possible frequency $b = \sigma_{\max} = n$.
We run an FSM algorithm with the frequency $\sigma = (a + b) / 2$ and set $a = \sigma$ if the mining process terminates earlier, and 
$b = \sigma$ otherwise,
and repeat the process until $a - b = 1$.
We can also use the early termination with the number of admissible subgraphs proposed in the incremental search above, which enables us to gain more efficiency and to determine whether the current frequency $\sigma$ is larger than the root.
This method could potentially reduce the number of frequencies to be examined. 

\begin{algorithm}[t]
 \caption{Bisection search (LEAP search)}
 \label{Algorithm:bisection}
 \begin{algorithmic}
  \STATE {\bfseries Input:} Datasets $\Gc$, $\Gc'$ and significance level $\alpha$
  \STATE {\bfseries Output:} All significant subgraphs
  \STATE $\sigma_{\min} \gets 1$
  \WHILE{$\psi(\sigma_{\min}) > \alpha$}
  \STATE $\sigma_{\min} \gets \sigma_{\min} + 1$
  \ENDWHILE
  \STATE $\sigma_{\max} \gets n$ \hspace*{10pt}\COMMENT{the maximum possible frequency}
  \STATE $\sigma_{\mathrm{rt}} \gets \lfloor (\sigma_{\min} + \sigma_{\max}) / 2 \rfloor$
  \REPEAT
  \STATE Run an FSM algorithm with the threshold $\sigma_{\mathrm{rt}}$
  \STATE with monitoring the number $m$ of frequent subgraphs
  \IF{$m > \alpha / \psi(\sigma_{\mathrm{rt}})$ while the process}
  \STATE Terminate the mining process
  \ELSE
  \STATE $\Hc(\sigma_{\mathrm{rt}}) \gets \{H \in \Hc \mid f(H) \ge \sigma_{\mathrm{rt}}\}$
  \ENDIF
  \IF{the mining process is terminated}
  \STATE $\sigma_{\min} \gets \sigma_{\mathrm{rt}}$
  \ELSE
  \STATE $\sigma_{\max} \gets \sigma_{\mathrm{rt}}$
  \ENDIF
  \STATE $\sigma_{\mathrm{rt}} \gets \lfloor (\sigma_{\min} + \sigma_{\max}) / 2 \rfloor$
  \UNTIL{$\sigma_{\max} - \sigma_{\min} = 1$}
  \IF{the last mining process was terminated}
  \STATE $\sigma_{\mathrm{rt}} \gets \sigma_{\max}$ \hspace*{10pt}\COMMENT{$\sigma_{\mathrm{rt}}$ is the root frequency}
  \ENDIF
  \STATE $\tau(\Hc) \gets \set{H \in \Hc | f(H) \ge \sigma_{\mathrm rt}}$
  \STATE \COMMENT{Testable hypotheses}
  \STATE $s(\Hc) \gets \set{H \in \tau(\Hc) | P\ \text{value of}\ H\ < \alpha / |\tau(\Hc)|}$
  \STATE Output $s(\Hc)$
 \end{algorithmic}
\end{algorithm}

\subsection{Effective number of tests}\label{subsec:effective}
Many subgraphs are expected to be highly correlated with each other due to combinatorial constraints on graphs such as subgraph-supergraph relationships~\cite{Ugander13}.
To exploit the dependence between subgraphs and further increase the power, we use the {\em effective number of tests}.
In the \v{S}id{\'a}k correction~\cite{Sidak67}, the significance level $\alpha'$ for each test is given as $1 - (1 - \alpha)^{1 / m}$ for $m$ {\em independent} tests.
This means that if we have $m$ tests and some of them are correlated, only $m_{\text{eff}} < m$ tests, defined by
\[
 m_{\text{eff}} := \frac{\log (1 - \alpha)}{\log (1 - \alpha')},
\]
are {\em effective} for controlling the FWER~\cite{Moskvina08}, hence $m_{\text{eff}}$ can be used as a reduced Bonferroni factor.
This $m_{\text{eff}}$ is called the effective number of tests and estimation methods, such as the Cheverud-Nyholt estimate~\cite{Nyholt04}, have been proposed in particular in statistical genetics.

We directly estimate the significance level $\alpha'$ for each test by random permutations of class labels, which gives the null distribution of independent subgraphs.
Although this method gives the optimal estimation of $m_{\text{eff}}$ in theory, its drawback is the high computational cost $O(mh)$ ($m = |\Hc|$ in our case), where $h$ is the number of iterations.
Here we overcome this drawback by considering only testable subgraphs.
Since we can ignore untestable hypotheses (subgraphs) for controlling the FWER, we apply the above permutation-based estimation to only testable subgraphs.
The complexity reduces to $O(|\tau(\Hc)|h)$, which is expected to be much cheaper than $O(|\Hc|h)$ if we can eliminate many untestable subgraphs.
We set the number of permutations to be 1,000 throughout the paper, which is recommended for $\alpha = 0.05$~\cite{Churchill94} and commonly used~\cite{Moskvina08}.

\section{Related Work}\label{sec:related}
The statistical significance of subgraph occurrence in networks has been investigated before, first in specific application domains, such as social networks~\cite{Wasserman94} and gene regulatory networks~\cite{Shen02}, and the formulation was later extended to general graphs~\cite{Arora14,He06,Milo02,Ranu09,Yan08}. 
In all of these studies, however, the significance is defined using a random database, that is, the $p$-value of a subgraph is the probability of its frequency being larger than the user-specified threshold under a certain distribution of graphs (or labels on graphs) and, to the best of our knowledge, no study directly detects subgraphs that are significantly associated with class memberships of graphs.
Moreover, our method overcomes the following three drawbacks of previous approaches:
(1) their $p$-values depend on the frequency threshold, which is often difficult to determine in practice, while our method requires only the significance level $\alpha$;
(2) their $p$-value computation requires a distribution of graphs, which is not trivial to estimate, while our method does not need to consider such a distribution and can still calculate the exact $p$-values;
(3) to the best of our knowledge, all previous studies did not consider the multiple testing problem, which leads to many false positives, while our method strictly controls the FWER.

Subgraph detection has also been intensively studied in {\em graph classification}, where subgraphs are used as {\em features} to describe graphs.
This means that each graph $G$ is represented as a feature vector in which each feature corresponds to another graph $H$ and the value is one if $H \sqsubseteq G$ and zero otherwise.
The general objective is to find informative subgraphs for discrimination to improve the accuracy of the subsequent classification, which can also be viewed as a supervised feature selection problem.
A number of methods have been proposed, for example, gBoost~\cite{Kudo05} and a Lasso-based method~\cite{Tsuda07}.
Note that, however, in classification we do not need to control the FWER (false positives) as long as we can build a good classifier,
while our ultimate goal in this paper is to detect key substructures for a better understanding of the target phenomenon and the FWER must be controlled to avoid false positives for further investigation in application domains.

Multiple (hypothesis) testing is a classical problem in statistics, with Bonferroni correction~\cite{Bonferroni36} being the most prominent correction technique. Since Bonferroni correction is known to be too conservative, other correction methods have been proposed, for instance, Holm's correction~\cite{Holm79}.
However, these methods also require the exact number of tests (subgraphs) for correction, which is highly expensive to compute in graph mining.
Another approach is to use random subsampling to estimate the correction factor~\cite{Dudoit03}, but this also needs high computational cost if the number of tests is massive.
Controlling the false discovery rate (FDR)~\cite{Benjamini95} is recently becoming popular as an alternative to the FWER, which leads to more power in multiple testing.
However, it also requires the exact number of tests and hence is also extremely expensive to compute.

\section{Experiments}\label{sec:experiment}
We examined our methods on real-world graph data and compared them to the brute-force approach (BF for short) and two state-of-the-art approaches (LAMP and LEAP) in our framework.
BF na{\"i}vely enumerates subgraphs occurring more than once to set the Bonferroni correction factor.
Notice that, with respect to assessing the quality of results, that is, the number of significant subgraphs, BF can be our only comparison partner,
since there exists no method for finding significant subgraphs while controlling the FWER by multiple testing correction.
On the efficiency side, we compare BF and our four search strategies, in which two of them (decremental LAMP search and bisection LEAP search) are the state-of-the-art.

As an FSM algorithm, we employ Gaston~\cite{Nijssen04} since it is reported to be one of the fastest FSM algorithms~\cite{Worlein05}.
We integrated our search strategies into Gaston, which are written in C++ and compiled with {\ttfamily gcc} 4.6.3.
The significance level $\alpha$ was always set to $0.05$ and a two-tailed test was used.
We repeated $1,000$ permutations to obtain the effective number of tests.
We used Ubuntu version 12.04.3 with a single 2.6 GHz AMD Opteron CPU and 512 GB of memory.
All experiments were performed in R 3.0.1.

\begin{table*}[t]
 \centering
 \caption{Statistics of datasets, where $|L(V)|$ and $|L(E)|$ denote the number of node and edge labels.}
 \label{Table:datasets}
   \begin{small}
   {\centering\tabcolsep = 1.5mm
   \begin{tabular}{lrrrrrrrrrrrr}
 \toprule
 Dataset & Size & \#positive & avg.$|V|$ & avg.$|E|$ & max$|V|$ & max$|E|$ & min$|V|$ & min$|E|$ & avg.deg & $|L(V)|$ & $|L(E)|$ \\ 
 \midrule
 PTC (MR) & 584 & 181 & 31.96 & 32.71 & 181 & 181 &   2 &   1 & 2.01 &   7 &   4 \\ 
 MUTAG & 188 & 125 & 17.93 & 39.59 &  28 &  66 &  10 &  20 & 4.38 &   7 &  11 \\ 
 ENZYMES & 600 & 300 & 32.63 & 62.14 & 126 & 149 &   2 &   1 & 3.86 &   3 &   1 \\ 
 D\&D & 1178 & 691 & 284.32 & 715.66 & 5748 & 14267 &  30 &  63 & 4.98 &  82 &   1 \\ 
 NCI1  & 4208 & 2104 & 60.12 & 62.72 & 462 & 468 &   4 &   3 & 2.08 &   8 &   4 \\ 
 NCI41 & 27965 & 1623 & 47.97 & 50.15 & 462 & 468 &   3 &   2 & 2.09 &   8 &   4 \\ 
 NCI167 & 80581 & 9615 & 39.70 & 41.05 & 482 & 478 &   2 &   1 & 2.06 &   8 &   4 \\
 NCI220 & 900 & 290 & 46.87 & 48.52 & 239 & 255 &   2 &   1 & 2.05 &   7 &   3 \\ 
 \bottomrule
   \end{tabular}
     }
   \end{small}
\end{table*}

\subsubsection*{Dataset}s
We used eight real-world graph datasets: PTC(MR), MUTAG, ENZYMES, D\&D, and four NCI datasets, where ENZYMES and D\&D are proteins and others are chemical compounds.
Statistics for these datasets are summarized in Table~\ref{Table:datasets}.
These datasets have been frequently used as benchmarks in previous studies \cite{Li12,Shervashidze11,Zhao11ICDM}.
They are labeled undirected graphs:
Graph nodes are labeled in all datasets and edges are also labeled except for ENZYMES and D\&D.

The PTC (Predictive Toxicology Challenge) dataset\footnote{\url{http://www.predictive-toxicology.org/ptc/}} contains data of 601 chemical compounds in total (including training and test sets),
which is originally designed for a prediction challenge of carcinogenic effects.
Graphs are classified according to their carcinogenicity assayed on rats and mice.
We assume that graphs labeled as CE, SE, or P as positive, and those of NE or N as negative, the same setting as in~\cite{Kong10,Zhao11ICDM}.
The dataset is divided into four overlapping subsets according to their animal models: male rats (MR), female rats (FR), male mice (MM), and female mice (FM).
We used only MR since the properties of other datasets are similar.

MUTAG~\cite{Debnath91} is a dataset of 188 mutagenic aromatic and heteroaromatic nitro compounds, which are classified into two classes of mutagenically active or inactive on the bacterium {\em Salmonella typhimurium}.

ENZYMES is a dataset of protein tertiary structures used in~\cite{Borgwardt05bio}, which consists of 600 enzymes, extracted from the BRENDA database~\cite{Schomburg04}.
Each enzyme is classified into one of six Enzyme Commission top level enzyme classes (EC1 to EC6).
We classified enzymes from EC1 to EC3 to one class, and from EC4 to EC6 to the other for our binary classification problem.

D\&D is a dataset of 1178 protein structures created by Dobson and Doig~\cite{Dobson03}, and they are classified into enzymes and non-enzymes.
As we can see in Table~1, the size of each graph in this dataset is relatively large compared to the other datasets%
\footnote{MUTAG, ENZYMES, and D\&D are obtained from \url{http://mlcb.is.tuebingen.mpg.de/Mitarbeiter/Nino/Graphkernels/data.zip}}.

NCI (National Cancer Institute) datasets contain data of chemical compounds that are classified according to their anti-cancer activity~\cite{Wale08}.
Datasets are numbered by their bioassay IDs.
NCI1 is balanced subsets, which is often used in the literature~\cite{Li12,Shervashidze11}, and the others are the full sets retrieved from the official website\footnote{\url{https://pubchem.ncbi.nlm.nih.gov/}}.

\subsubsection*{Effectiveness}
First we compare the Bonferroni correction factors and our reduced correction factors, that is, the number of testable subgraphs $|\tau(\Hc)|$ and that of effective subgraphs $m_{\text{eff}}$, and
evaluate the improvement of our method in terms of the power for detecting significant subgraphs and the empirical FWERs obtained from 10,000 permutations of class labels.
In each dataset, we varied the upper bound of the subgraph size from $4$ to $16$ and without size bound (``Limitless'').

\begin{figure}[t]%
 \centering
 \includegraphics[width=.8\linewidth]{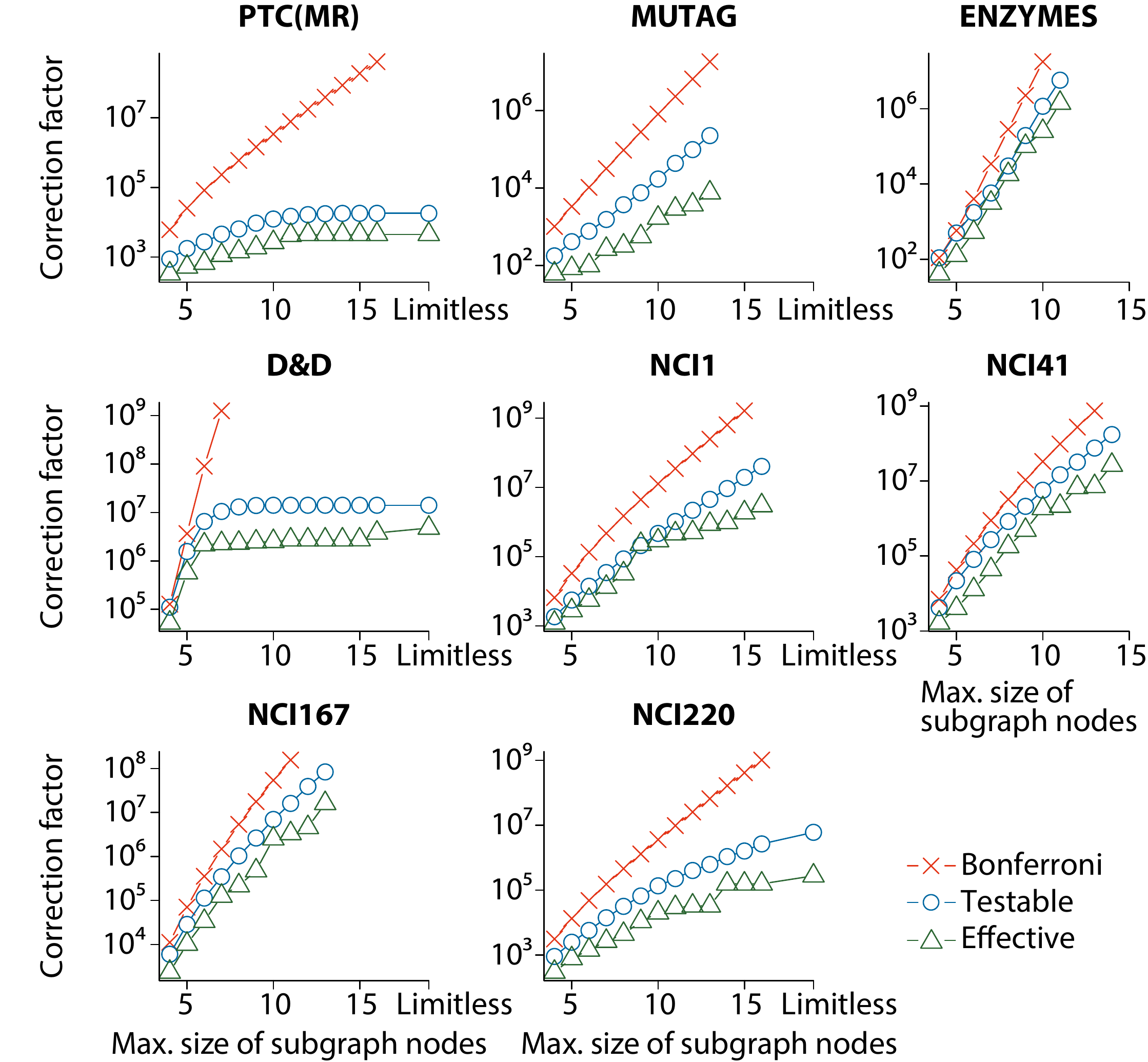}
 \caption{The Bonferroni correction factor $|\Hc|$ (red cross marks), the number of testable subgraphs $|\tau(\Hc)|$ (blue circles), and the effective number of tests $m_{\text{eff}}$ (green triangles).  Note that the $y$-axis has a logarithmic scale.}
 \label{Figure:factor}
\end{figure}

\begin{figure}[t]%
 \centering
 \includegraphics[width=.8\linewidth]{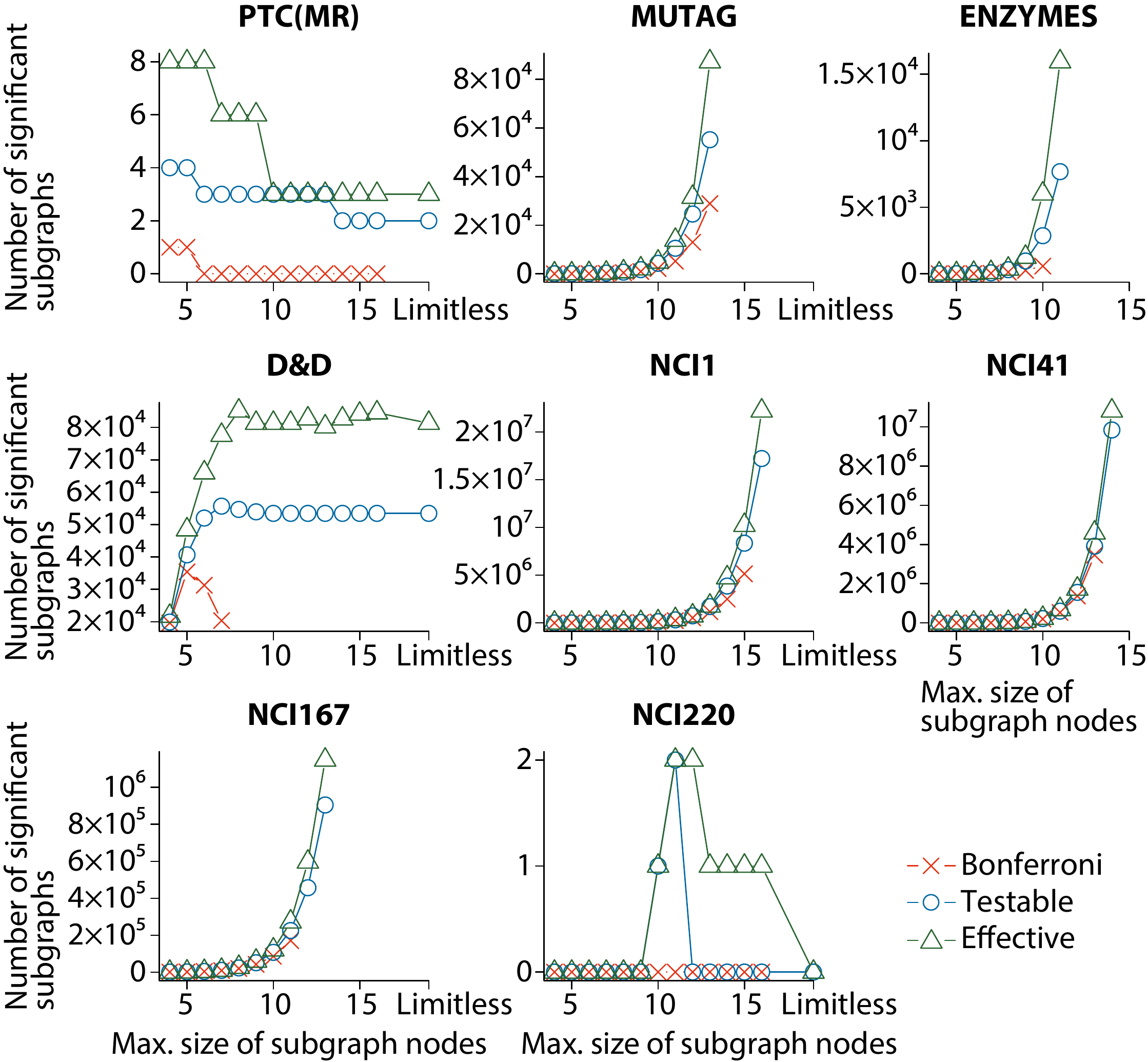}
 \caption{The number of significant subgraphs discovered with the Bonferroni correction (red cross marks) and our method with the testability criterion (blue circles) and the effective number of tests (green triangles).}
 \label{Figure:signum}
\end{figure}

\begin{figure}[t]%
 \centering
 \includegraphics[width=.8\linewidth]{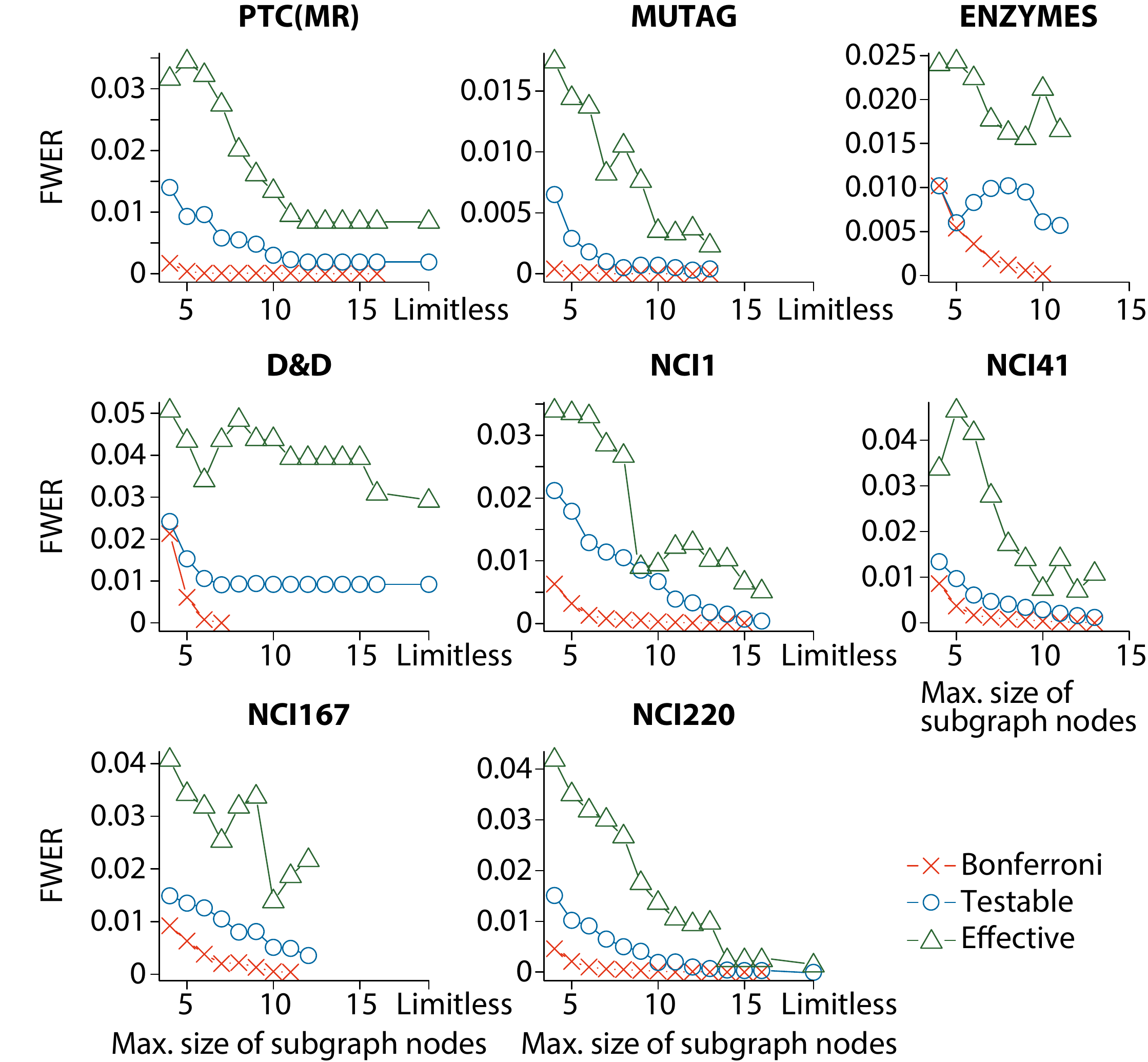}
 \caption{The empirical FWERs with 10,000 permutations of class labels with the Bonferroni correction (red cross marks), our method with the testability criterion (blue circles) and the effective number of tests (green triangles).}
 \label{Figure:FWER}
\end{figure}

The resulting correction factors are plotted in Figure~\ref{Figure:factor} and the numbers of significant subgraphs we detected and the empirical FWERs are shown in Figures~\ref{Figure:signum} and~\ref{Figure:FWER}, respectively.
There are some missing values in the plots, in particular results of the Bonferroni factor (red cross marks), due to a huge amount of computation time.
These plots clearly show that, in all datasets, our correction factor is much smaller than the Bonferroni factor and the difference between them becomes larger as the maximum subgraph size increases.
In particular in PTC(MR) and D\&D, our factors (blue circles and green triangles) become stable in large maximum subgraph sizes while the Bonferroni factors increase exponentially.
The reason might be that most of large subgraphs become untestable because they tend to have small frequencies in general.
Moreover, we can confirm that in all datasets correction factors are further reduced using the effective number of tests.
This is because many subgraphs are highly correlated with each other due to combinatorial constraints of graphs~\cite{Ugander13}.

\begin{figure}[t]%
 \centering
 \includegraphics[width=\linewidth]{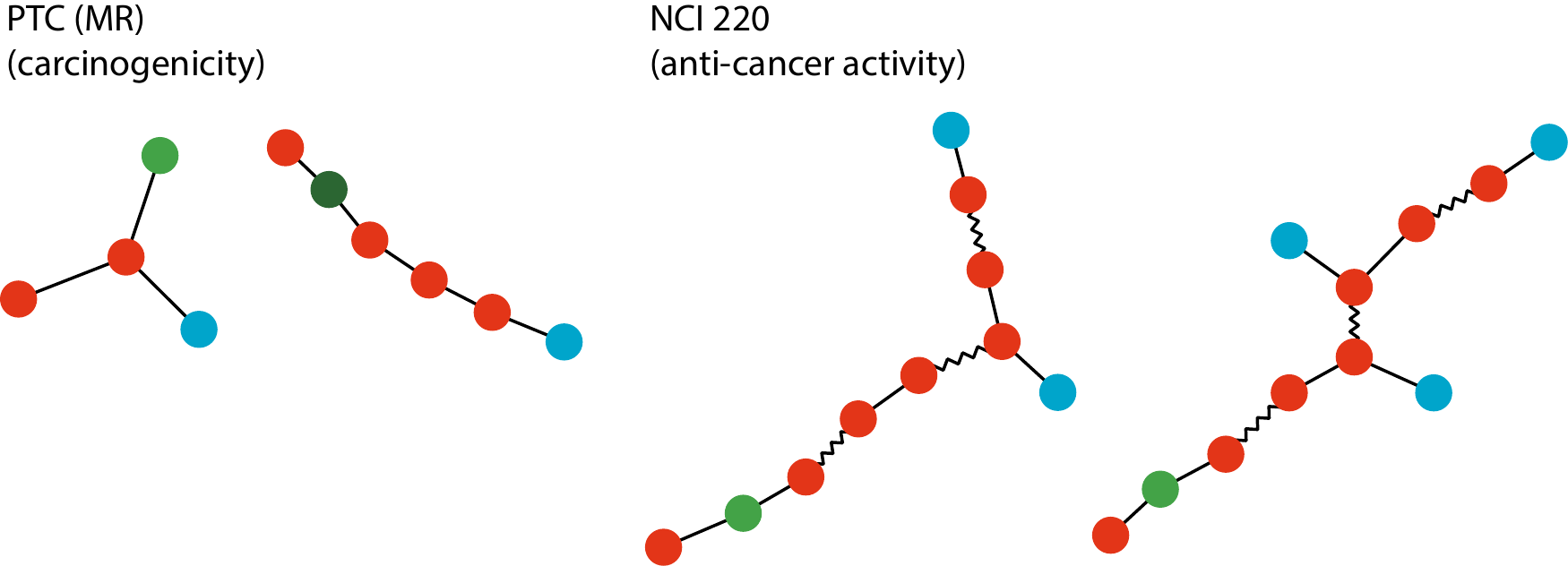}
 \caption{Four examples of significant subgraphs on PTC(MR) (left) and NCI220 (right) that are detected by our method using the testability but are missed by the standard Bonferroni factor. Different colors (resp. shapes) of vertices (resp. edges) mean different labels of them.}
 \label{Figure:example}
\end{figure}

In terms of the number of significant subgraphs (Figure~\ref{Figure:signum}), we can find more subgraphs because of the reduced correction factor across our datasets. On several datasets the effect is dramatic, such as MUTAG, ENZYMES or D\&D, where our methods find thousands of significant subgraphs missed by the standard Bonferroni correction. Examples are shown in Figure~\ref{Figure:example}.
In PTC(MR), one cannot find any significant subgraphs by the Bonferroni correction when the maximum subgraph size is larger than $6$,
but one can detect $2$ to $4$ (testable) or $3$ to $8$ (effective) significant subgraphs using our factors.
Moreover, the number of significant subgraphs in the Bonferroni factor rapidly decreases in the D\&D dataset as the maximum subgraph size increases, while numbers are stable in our methods even if the maximum subgraph size is unlimited. Since it is often difficult to appropriately upper bound the subgraph size beforehand in practice,
this is another advantage in practical applications.
In NCI220 the number of significant subgraphs exhibits an interesting behavior, that is, significant subgraphs are detected only if the maximum subgraph size is $10$ or $11$ (testable) and from $10$ to $16$ (effective).
The reason is that the size of these significant subgraphs is $10$ or $11$ and we cannot detect them if the maximum subgraph size is smaller than that.
Furthermore, these subgraphs are no longer significant if the maximum subgraph size becomes larger due to the increase of the correction factor.

We can also confirm the higher statistical power from the empirical FWERs (Figure~\ref{Figure:FWER}).
Note that the FWER should be $\alpha = 0.05$ in the best case, and the correction factor is too large if the FWER is smaller than $\alpha$.
By reducing the correction factor with the testability criterion and the effective number of tests, the FWERs get closer to $\alpha = 0.05$.

\begin{figure}[t]%
 \centering
 \includegraphics[width=.8\linewidth]{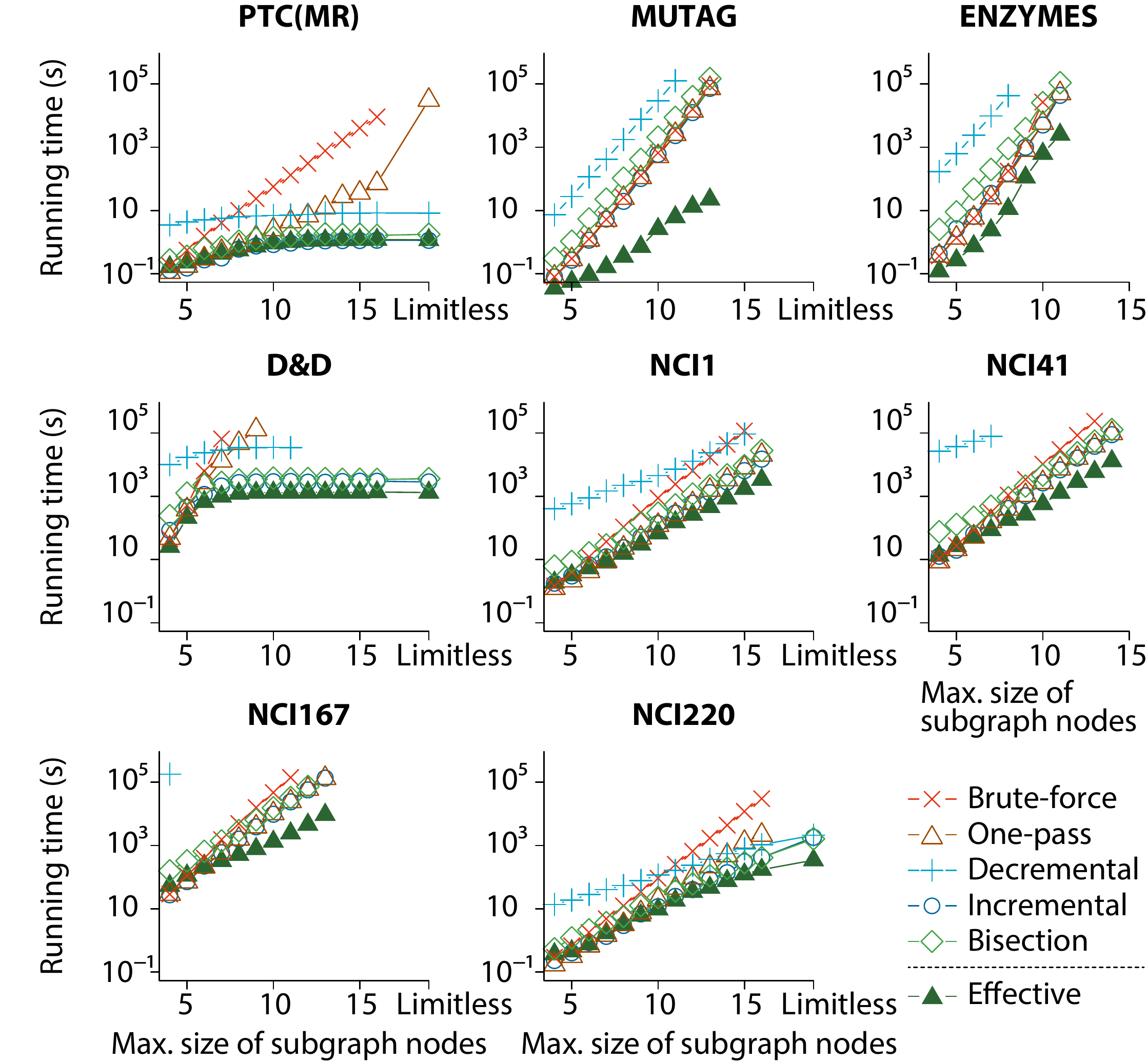}
 \caption{Running time (seconds). Note that the $y$-axis is in logarithmic scale.}
 \label{Figure:runtime}
\end{figure}%

\subsubsection*{Efficiency}
Next we analyze the efficiency of our strategies compared to BF and the state-of-the-art (LAMP and LEAP).
The resulting running times are plotted in Figure~\ref{Figure:runtime} and are summarized in Table~\ref{Table:runtime}
as RMSD (root mean square deviation) to the best (fastest) running time on each dataset and for each maximum subgraph size.
In addition, we also plot the running time of computing the effective number of testable subgraphs by using 1,000 permutations in Figure~\ref{Figure:runtime}.

The results clearly show that all four searches using the testability criterion are faster than BF on average.
This means that reducing the number of subgraph candidates using the testability of them contributes not only to the effectiveness in terms of finding significant subgraphs but also the efficiency of the whole process.
Furthermore, our new incremental search is one to two orders of magnitude faster than the other state-of-the-art search strategies (decremental LAMP and bisection LEAP) and more than two orders of magnitude faster than the one-pass search and BF on average.
In contrast, the decremental LAMP search is slow, with its speed being similar to the one-pass search on average, and it is often even slower than BF.
The reason is that in practice the root frequency $\sigma_{\mathrm{rt}}$ is relatively small (around $20$, see Table~\ref{Table:root}) and hence the decremental search needs to repeat an FSM algorithm many times until reaching this frequency.
This is also the reason for the efficiency of the incremental search as it can quickly find the root frequency.
Although the bisection LEAP search is faster than the decremental and the one-pass search on average, it is slower than the incremental search.
The reason is the same as in the discussion above, that is, the root frequency is usually small and it tends to repeat subgraph mining with high frequencies.

The running time for computing the effective number of tests is faster than the above-mentioned search of testable subgraphs in most cases.
This means that the testability criterion also contributes to the efficiency of computing the effective number of tests and makes it feasible within a reasonable time.

\begin{table}[t]
 \centering
 \caption{Root frequencies $\sigma_{\text{rt}}$ for each dataset and each maximum size of subgraph nodes.
 ``---'' means that computation did not finished and the root frequency is not confirmed.}
 \label{Table:root}
 \begin{tabular}{lrrrrrrrrrrrrrr}
  \toprule
  & \multicolumn{14}{c}{Maximum size of subgraph nodes}\\
  Dataset & 4 & 5 & 6 & 7 & 8 & 9 & 10 & 11 & 12 & 13 & 14 & 15 & 16 & Limitless \\ 
  \midrule
  PTC(MR) & 9 & 9 & 10 & 10 & 11 & 11 & 11 & 11 & 11 & 11 & 11 & 11 & 11 & 11 \\ 
  MUTAG & 8 & 8 & 9 & 10 & 10 & 11 & 12 & 12 & 13 & 14 & --- & --- & --- & --- \\ 
  ENZYMES & 11 & 14 & 15 & 17 & 19 & 22 & 24 & 27 & --- & --- & --- & --- & --- & --- \\ 
  D\&D & 17 & 20 & 21 & 22 & 22 & 22 & 22 & 22 & 22 & 22 & 22 & 22 & 22 & 22 \\ 
  NCI1 & 16 & 17 & 19 & 20 & 21 & 22 & 24 & 25 & 26 & 27 & 28 & 29 & 30 & --- \\ 
  NCI41 & 5 & 5 & 6 & 6 & 6 & 7 & 7 & 7 & 8 & 8 & 8 & --- & --- & --- \\ 
  NCI167 & 6 & 7 & 7 & 8 & 8 & 9 & 9 & 10 & 10 & 11 & --- & --- & --- & --- \\ 
  NCI220 & 9 & 10 & 11 & 11 & 12 & 13 & 13 & 14 & 14 & 15 & 15 & 16 & 16 & 18 \\ 
   \bottomrule
\end{tabular}
\end{table}

\section{Conclusion}\label{sec:conclusion}
In this paper, we have presented a solution for finding subgraphs that are statistically significantly enriched in one class of graphs but not another.
The difficulty of the problem stems from the two facts that (1) one has to consider an enormous search space of candidate subgraphs
and that (2) one has to correct the significance level for multiple testing to control the FWER, as one tests a large number of candidate subgraphs simultaneously.
The first problem leads to enormous computational runtime problems, the second one to a loss in the statistical power to detect significant subgraphs.

We have shown that the problem can be exactly and efficiently solved by considering only \textit{testable} subgraphs, which include all significant subgraphs and dramatically reduce the number of tests performed, thereby leading to a gain in statistical power.
Moreover, we can further increase the power using the effective number of tests, which reduces the correction factor according to the dependence between subgraphs.
We have presented several search strategies that use frequent subgraph mining algorithms to efficiently retrieve the set of testable subgraphs.
Experimental results show that our method finds significant subgraphs with higher speed and higher statistical power than any state-of-the-art approach.
This result promises to open the door to many interesting applications in chemoinformatics, structural biology and personalized medicine.

We also believe that our approach lays the foundation for follow-up studies in several important directions:
developing and integrating other approaches which exploit the dependence between tests~\cite{Zhang08},
considering other types of structured data such as strings, and
summarizing the solution set of significant subgraphs, which sometimes grows extremely large.

\begin{table*}[t]
 \centering
 \caption{RMSD (root mean square deviation) of running time (seconds) in Figure~\ref{Figure:runtime} to the best (fastest) running time on all datasets and maximum subgraph sizes. This measure rewards methods that are always close to the fastest running time on each dataset and each maximum subgraph size.}
 \label{Table:runtime}
\begin{tabular}{cccccc}
 \toprule
 Brute-force (BF) & One-pass & Decremental (LAMP) & Incremental & Bisection (LEAP)\\
 \midrule
  $6.994 \times 10^4$ & $2.635 \times 10^4$ & $2.410 \times 10^4$ & $\mathbf{1.230 \times 10^2}$  & $9.554 \times 10^3$\\
 \bottomrule
\end{tabular}
\end{table*}

\section*{Acknowledgments}
This work was funded in part by a Grant-in-Aid for Scientific Research (Research Activity Start-up) 26880013 (MS), the SNSF Starting Grant ``Significant Pattern Mining'' (KMB), the Alfried Krupp von Bohlen und Halbach-Stiftung (KMB), and the Marie Curie Initial Training Network MLPM2012, Grant No. 316861. (FLL, KMB).

\end{document}